\documentclass[11pt]{article}

\usepackage{cite}
\usepackage{amssymb,amsmath,amsfonts}
\usepackage{latexsym}
\usepackage{mathrsfs}
\usepackage{hyperref} 
\usepackage{epsfig,verbatim}
\usepackage{graphics}
\usepackage{comment}
\usepackage{subfigure,slashed}

\usepackage{xcolor}
\newcommand{\new}[1]{\color{black}{#1}\color{black}}

\def\k{{\bf k}}

\def\R{{\mathcal R}}
\def\P{{\mathcal P}}

\def\tk{{\tilde{k}}}

\def\bea{\begin{eqnarray}}
\def\eea{\end{eqnarray}}
\def\be{\begin{equation}}
\def\ee{\end{equation}}
\def\ba{\begin{array}}
\def\ea{\end{array}}
\def\nn{\nonumber}

\begin{document}

\setlength\arraycolsep{2pt}

\renewcommand{\theequation}{\arabic{section}.\arabic{equation}}
\setcounter{page}{1}

\begin{titlepage}

\begin{center}

\vskip 1.0 cm

{\LARGE  \bf Consistency relations for sharp inflationary non-Gaussian features}

\vskip 1.0cm

{\large
Sander Mooij${}^{a}$, Gonzalo A. Palma${}^{a}$, Grigoris Panotopoulos${}^{a}$ and Alex Soto${}^{b}$
}

\vskip 0.5cm

{\it
${}^{a}$Departamento de F\'{i}sica, Facultad de Ciencias F\'{i}sicas y Matem\'{a}ticas, Universidad de Chile\\ 
\mbox{Blanco Encalada 2008, Santiago, Chile} \\
\mbox{${}^{b}$Departamento de F\'{i}sica, Facultad de Ciencias, Universidad de Chile}\\ 
\mbox{Las Palmeras 3425, \~{N}u\~{n}oa, Santiago, Chile} \\
}

\vskip 1.5cm

\end{center}

\begin{abstract}

If cosmic inflation suffered tiny time-dependent deviations from the slow-roll regime, these would induce the existence of small scale-dependent features imprinted in the primordial spectra, with their shapes and sizes revealing information about the physics that produced them. Small sharp features could be suppressed at the level of the two-point correlation function, making them undetectable in the power spectrum, but could be amplified at the level of the three-point correlation function, offering us a window of opportunity to uncover them in the non-Gaussian bispectrum. In this article, we show that sharp features may be analyzed using only data coming from the three point correlation function parametrizing primordial non-Gaussianity. More precisely, we show that if features appear in a particular non-Gaussian triangle configuration (\emph{e.g.} equilateral, folded, squeezed), these must reappear in every other configuration according to a specific relation allowing us to correlate features across the non-Gaussian bispectrum. As a result, we offer a method to study scale-dependent features generated during inflation that depends only on data coming from measurements of non-Gaussianity, allowing us to omit data from the power spectrum.

\end{abstract}

\end{titlepage}

\newpage

\section{Introduction} 
\setcounter{equation}{0}

Despite of the fact that the simplest models of cosmic inflation \cite{Guth:1980zm,Linde:1981mu,Albrecht:1982wi, Starobinsky:1980te} predict primordial curvature perturbations distributed according to a nearly Gaussian statistic parametrized by a scale independent set of spectra \cite{Mukhanov:1981xt,Maldacena:2002vr}, there are good reasons to consider scenarios in which scale-dependent features are generated during inflation. If cosmic inflation experienced tiny time-dependent deviations from the slow-roll regime, these would show up in the primordial spectra in the form of small scale-dependent features, with their shapes and sizes revealing important information about the physics that produced such deviations in the first place. Moreover, these features would consistently appear in every $n$-point correlation function, leading to correlated features in the primordial spectra. 

Indeed, during the past few years, several works \cite{Hotchkiss:2009pj,Achucarro:2010da,Adshead:2011jq,Park:2012rh, Achucarro:2012fd,Saito:2013aqa,Noumi:2013cfa,Achucarro:2013cva,Miranda:2013wxa,Achucarro:2014msa,Gong:2014spa,Mizuno:2014jja,Ashoorioon:2014yua,Hu:2014hra,Fergusson:2014hya,Munchmeyer:2014cca,Palma:2014hra,Fergusson:2014tza,Gariazzo:2015qea,Chluba:2015bqa,Mooij:2015cxa,Appleby:2015bpw,Gao:2015aba,Cai:2015xla,Meerburg:2015owa} have emphasized the fact that if features are present in the power spectrum $\mathcal P(k)$, they should consistently reappear in the bispectrum $B(k_1, k_2 , k_3)$, and any other higher correlation function. Specifically, it is possible to deduce a general expression relating features appearing in the $f_{\rm NL}$-function (parametrizing departures from pure Gaussianity\footnote{In this work we use a version of the $f_{NL}$-parameter that is useful to parametrize non-Gaussianity at different triangle configurations. The definition is provided in eq.~(\ref{fnl}).}) with those appearing in the power spectrum $\P$, given by
\be
f_{NL} (k_1 , k_2 , k_3) = \left[ f_2  \frac{d^2}{d \ln k^2} \frac{\Delta \mathcal P }{\mathcal P_0} (k)  +  f_1 \frac{d}{d \ln k}  \frac{\Delta \mathcal P}{\mathcal P_0} (k)  +  f_0 \, \frac{\Delta \mathcal P }{\mathcal P_0} (k) \right]_{k = (k_1 + k_2 + k_3)/2} , \label{general_expression}
\ee
where $\Delta \mathcal P \equiv \mathcal P - \mathcal P_0$ (with $\mathcal P_0$ corresponding to the featureless power spectrum). In the previous expression, the functions $f_i \equiv f_i (k_1, k_2 , k_3)$ represent known functions\footnote{Please see refs.~\cite{Achucarro:2012fd, Palma:2014hra, Mooij:2015cxa} for explicit expressions valid in different circumstances, depending on the source that generated the time-deviations from the slow-roll regime.} of the triangle configuration determined by the scales $k_1$, $k_2$ and $k_3$, but that are scale independent ({\it i.e.} they are invariant under rescalements $k_i \to k_i' = \gamma \,  k_i$). Equation~(\ref{general_expression}) links features appearing in the power spectrum evaluated at the given scale $k$ with features in the bispectrum evaluated at the 2-dimensional surface given by $k_1 + k_2 + k_3 = 2 k$.  The relative values of the scales $k_1$, $k_2$ and $k_3$ determine the triangle configuration in momentum space, whereas $k_1 + k_2 + k_3$ gives us the size of the triangle. In the case of the squeezed configuration, where one of the three momenta is much smaller than the other two ({\it e.g.} $k_3 \ll k_1 , k_2$), one finds that $f_2, f_0 \ll 1$, and so we recover the non-Gaussian consistency relation~\cite{Maldacena:2002vr,Creminelli:2004yq}, where $f_1 \to - 5 / 12$.

The correlation shown in eq.~(\ref{general_expression}) was first derived in ref.~\cite{Achucarro:2012fd}, where features in the spectra were studied as a consequence of time-variations of the sound speed $c_s$ of primordial perturbations away from the canonical value $c_s = 1$. It was later deduced in ref.~\cite{Palma:2014hra} for the case in which features are generated by time-variations of the Hubble expansion rate $H$ alone (implying time-variations of the slow-roll parameters $\epsilon$ and $\eta$). Then, it was generalized in ref.~\cite{Mooij:2015cxa} to the case in which features are generated when both classes of time-variations happen simultaneously. In the latter case, the coefficients $f_2$, $f_1$ and $f_0$ depend on a single parameter that quantifies the mixing between both types of time variations (sound speed $c_s$ vs Hubble expansion rate $H$). These coefficients constitute predictions from inflation, and in principle may be tested by Cosmic Microwave Background (CMB) and Large Scale Structure (LSS) observations. 

\subsection{Main idea}

If the small time-dependent deviations from the slow-roll regime happen at a fast rate compared to the inverse of the Hubble expansion rate $H$ during inflation, then the features in the power spectrum $\Delta \mathcal P$ become sharp, in the sense that higher order derivatives with respect to $\ln k$ dominate.\footnote{We note that eq.~(\ref{general_expression}) in principle contains higher derivatives of $\Delta \P$ as well. However, these are slow-roll suppressed \cite{Mooij:2015cxa}. The highest unsuppressed term is the one containing the second derivative of the power spectrum.} This implies that in the sharp feature limit, and as long as we examine the bispectrum away from the squeezed configuration, the correlation of eq.~(\ref{general_expression}) becomes dominated by the term proportional to $f_2$, leading to a simpler version of it, given by
\be
f_{\rm NL} (k_1 , k_2 , k_3) =  \beta (k_1 , k_2 , k_3)  \frac{d^2}{d \ln k^2} \frac{\Delta \mathcal P }{\mathcal P_0} (k)   \bigg|_{k = (k_1 + k_2 + k_3)/2} , \label{general_expression_sharp}
\ee
where, for simplicity, we have substituted $\beta \equiv f_2$. As a consequence, if features are present in the power spectrum but with a small amplitude ({\it i.e.} $\Delta \mathcal P / \mathcal P_0 \ll 1$), their sharpness will nevertheless enhance the amplitude of features appearing in the bispectrum $f_{\rm NL} (k_1 , k_2 , k_3)$. \new{In this way, one could even imagine to measure features in the bispectrum before observing them in the power spectrum. However, most likely this will not be the case. As first pointed out in \cite{Behbahani:2011it}, and then worked out much further in \cite{Adshead:2013zfa,Adshead:2014sga}, by the very nature of the EFT framework, the signal-to-noise ratio is in general larger in the power spectrum than in the bispectrum. Typically, in situations where bispectrum features are more significant than power spectrum ones, these take place at energies close to (or beyond) the cut-off, i.e. energy scales where one stops to trust the EFT approach. (See however \cite{Behbahani:2012be} for an instructive counterexample, in which collective symmetry breaking ensures that breaking of scale invariance takes place at the same order in perturbation theory for any $N$-point function.) 

At any rate, bispectrum features may become large in the sharp feature limit, and will manifest itself in different shapes. This observation motivates us to consider the development of additional theoretical tools allowing us to analyze the presence of features in the bispectrum alone, without the need of using information coming from features in the power spectrum.}

%Thus, even if features are not measurable at the level of the power spectrum, it is still conceivable that they could be observed and analyzed in the bispectrum. This observation motivates us to consider the development of additional theoretical tools allowing us to analyze the presence of features in the bispectrum alone, without the need of using information coming from features in the power spectrum, which, after all, may never be measured. 

In fact, eq.~(\ref{general_expression_sharp}) already gives us a hint of how one could study features in the bispectrum alone. For example, if we examine the equilateral configuration, where $k_1 = k_2 = k_3 = K/3$ (in such a way that $k_1 + k_2 + k_3 = K$) we obtain from eq.~(\ref{general_expression_sharp}) that:
\be
f_{\rm NL} (K/3 , K/3 , K/3) = \left[ \beta^{(\rm{eq})}  \frac{d^2}{d \ln k^2} \frac{\Delta \mathcal P }{\mathcal P_0} (k)  \right]_{k = K/2} . \label{equilateral_limit}
\ee
On the other hand, in the folded configuration, where $k_1 = k_2 = K/4$ and $k_3 = K/2$ (or cyclic permutations of $k_1$, $k_2$ and $k_3$) we deduce from eq.~(\ref{general_expression_sharp}) that:
\be
f_{\rm NL} (K/4 , K / 4,  K/2) = \left[ \beta^{(\rm{fold})}   \frac{d^2}{d \ln k^2} \frac{\Delta \mathcal P }{\mathcal P_0} (k)  \right]_{k = K/2} . \label{folded_limit}
\ee
Then, combining these two expressions we are able to deduce a relation linking features appearing in the equilateral configuration with those appearing in the folded configuration of the bispectrum. The result is simply given by:
\be
f_{\rm NL} (K/3 , K/3 , K/3) = \left( \frac{ \beta^{(\rm{eq})}}{ \beta^{(\rm{fold})}}  \right)  f_{\rm NL} (K/4 , K / 4,  K/2) . \label{main_idea}
\ee
Here, the ratio $\beta^{(\rm{eq})} / \beta^{(\rm{fold})}$ is a number that is independent of the scale $K$, but that depends on the specific type of time-deviation from slow-roll that generated the feature during inflation. Equation~(\ref{main_idea}) relates features appearing in different triangle configurations with a common scale $k_0$ (which characterizes the size of the triangle). Thus, eq~(\ref{main_idea}) gives us valuable information about how features appear in the bispectrum across different triangle configurations, and may be extended to a relation linking a specific configuration with any other desired configuration.

The aim of this article is to deduce an expression that generalizes eq.~(\ref{main_idea}), and that correlates features at different triangle configurations of the bispectrum. Our main result is given by eq.~(\ref{FINres2}), which constitutes the desired generalization of eq.~(\ref{main_idea}). In principle, that expression follows directly from eq.~(\ref{general_expression}), in the same way eq.~(\ref{main_idea}) was deduced. Nevertheless, in this paper we want to formally show that one can indeed compute the correlation between two bispectrum configurations without ever considering the associated feature in the power spectrum. To deduce it, we will use the effective field theory of inflation formalism to parametrize the interactions leading to the appearance of features in the  non-Gaussian bispectrum. 

Our article is organized as follows: in Section~\ref{sec:pre} we briefly cover the necessary background material developed in refs.~\cite{Achucarro:2012fd, Palma:2014hra, Mooij:2015cxa} necessary to study features in the primordial spectra. Then, in Section~\ref{corrgen} we compute the correlation between two generally (but not squeezed)  triangle configurations in momentum space. The inclusion of the squeezed limit is dealt with in Section~\ref{corrsq}. In the Appendix, we repeat the computations of Sections~\ref{corrgen} and \ref{corrsq} for a different parametrization of the momentum triangles, that some readers may find useful.

\subsection{Notation and conventions}

Before commencing the main part of our work, let us list our notation and conventions. We shall work with a background space-time metric given by the standard Friedman-Robertson-Walker metric of the form:
\be
ds^2 = - dt^2 + a^2(t) d {\bf x}^2,
\ee
where ${\bf x}$ represent comoving coordinates and $a(t)$ is the scale factor describing the expansion of spatially flat slices. The Hubble expansion parameter is given by $H = \dot a / a$, where the dot represents derivatives with respect to cosmic time $t$. We will also work with conformal time $\tau$, which is related to cosmic time trough the relation $d\tau = dt / a$. We will reserve primes to distinguish derivatives with respect to $\tau$.

\setcounter{equation}{0}
\section{Preliminaries} \label{sec:pre}

Our starting point is the effective field theory of inflation formalism \cite{Cheung:2007st,Senatore:2010wk}. Here, the evolution of primordial curvature perturbations is parametrized by two families of parameters. First, we have the standard slow-roll parameters which parametrize the evolution of the quasi-de Sitter spacetime described by the Hubble expansion rate $H$. These are given by:
\be
\epsilon = - \frac{\dot H}{H^2} , \qquad \eta = \frac{\dot \epsilon}{\epsilon H} .
\ee
The second class of quantities parametrizes deviations from canonical inflation. The most important parameter performing this task is the sound speed $c_s$ at which primordial perturbations propagate during inflation. When the action describing perturbations during inflation is written in terms of the curvature perturbation in co-moving gauge, $\mathcal R$, the parameter $c_s$ appears at different orders in the perturbative expansion. The total action $S$ describing the evolution of $\mathcal R$, up to cubic order, is found to be:
\be
S =  \int \! d^4 x \,  a^3 \epsilon \left[ \frac{1}{c_s^2} \dot {\mathcal{R}}^2 - \frac{1}{a^2}(\nabla \mathcal{R})^2  \right] + S^{(3)} ,
\ee
where the cubic contribution $S^{(3)}$ is given by~\cite{Burrage:2011hd}
\bea
S^{(3)} &=&  \int \! d^4 x \,  a^3 \epsilon \bigg[ 
\frac{1}{c_s^4} \left[ 3 (c_s^2 - 1) + \epsilon - \eta \right] \R \dot \R^2 + \frac{1}{c_s^2 a^2} \left( (1 - c_s^2) + \eta + \epsilon -  \frac{2 \dot c_s}{H c_s}  \right) \R (\nabla \R)^2  \nn \\
&&  + \frac{1}{H }  \left(\frac{1 - c_s^2}{c_s^4}  - \frac{ 2 \lambda }{ \epsilon H^2} \right)  \dot \R^3  + \frac{1}{4 a^4} (\partial \chi)^2 \nabla^2 \R - \frac{4 - \epsilon}{2 \epsilon a^4} \nabla^2 \chi \partial^i  \R \partial_i \chi + \frac{f }{\epsilon a^3} \frac{\delta S^{(2)}}{\delta \R}
 \bigg] , \label{S-R-cubic}
\eea
where $\chi$ is given by the constraint equation $\nabla^{2} \chi = a^2 \epsilon \dot \R /  c_s^{2}$. In addition, the parameter $\lambda$ in eq.~(\ref{S-R-cubic}) parametrizes the strength of the operator $\dot \R^3$, and it is usually found to depend on $c_s$ according to a relation determined by the specific model in question. Finally, the quantity $f$ multiplying the linear classical equation of motion $\delta S^{(2)} / \delta \R$ is a given quadratic function of $\R$, whose specific form will turn out to be irrelevant for the present discussion.

In this work we shall analyze small departures of the sound speed from the canonical background value $c_s = 1$. For this reason, it is useful to define the parameter $\theta$ given by:
\be
\theta \equiv 1 - c_s^2 .
\ee
By definition, one has $\theta \geq 0$. In addition, we are interested in rapid variations of the parameters $\epsilon$, $\eta$ and $c_s$ (and any other parameters depending on them). To be specific, by rapid variations we mean that the following hierarchical relation applies
\be
\Big| \tau \frac{d A}{d\tau} \Big| \gg | A |  , \label{hierarchy}
\ee
where $\tau$ is conformal time and the quantity $A$ can be either $\epsilon$, $\eta$ or $\theta$. If a quantity $A$ respects this relation, then its time variation is characterized by a timescale much smaller than $H^{-1}$. In fact, in our previous work \cite{Mooij:2015cxa} we proposed a relation linking both $\eta$ and $\theta$, given by
\be
\eta = -\frac{\alpha}{2} \tau \theta' ,\label{eta-theta-relation}
\ee
where $\alpha$ is a slowly varying quantity (a constant for all practical purposes) that parametrizes the specific class of model of inflation in which these rapid variations happen. While we have not found a proof of the general validity of this relation, we have verified that it is always satisfied in a large variety of non-canonical models of inflation, as long as $\theta \ll 1$ and the hierarchy of eq.~(\ref{hierarchy}) is satisfied. The parameter $\alpha$ can indeed be approximated to a constant, its numerical value depending on the parameter input of the model. Moving on with the discussion, after assuming that $\epsilon \ll 1$ and $\theta \ll 1$ and that these quantities respect a hierarchy of the form given in eq.~(\ref{hierarchy}), the cubic contribution to the action becomes:
\bea
S^{(3)}_{\rm int} = - \int \! d^4 x \,  a^3 \epsilon_0  \bigg\{   (3 \theta + \eta) \R \dot \R^2 + \frac{1}{a^2} (  \tau \theta' - \eta)  \R (\nabla \R)^2   \bigg\} .  \label{final-S3}
\eea
An important additional assumption that allowed us to arrive at eq.~(\ref{final-S3}) is that the coefficient in front of the operator $\dot {\mathcal R}^3$ (containing $\lambda$) is proportional to $(1 - c_s^2)^2$. This is in fact true for every known single field effective field theory representation of non-canonical models of inflation. Moreover, in ref.~\cite{Baumann:2015nta} it was recently conjectured that every operator (in the effective field theory of inflation expansion) parametrizing departures from non-canonical models of inflation must be proportional to powers of $1 - c_s^2$. This assumption implies that this term is subleading with respect to those that were preserved in the final expression for the cubic contribution to the action.

Our main goal is to compute the effect of sharp features on the three point correlation function. To accomplish this, it is convenient to use the in-in formalism to compute $n$-point correlation functions of $\mathcal R ({\bf x}, t )$ at the end of inflation. In particular, the three point correlation function $B (\k_1 , \k_2 , \k_3 )$  is defined in the following way
\be
 \langle \hat \R_{\k_1}  \hat \R_{\k_2}  \hat \R_{\k_3}  \rangle = (2 \pi)^3 \delta (\k_1 + \k_2 + \k_3 ) B (\k_1 , \k_2 , \k_3 ) ,
\ee
where $\hat \R_{\k_1}$ are the curvature perturbations in Fourier space evaluated at the end of inflation. To characterize features, the bispectrum can be written as $B=B_0+\Delta B$, where $B_0$ is the featureless part and $\Delta B$ represents the part that contains features. By using the in-in formalism to compute $n$-point correlation functions during inflation, it is straightforward to derive that $\Delta B$ is given by
\bea
\Delta B (\k_1, \k_2, \k_3) \! = \!   \frac{2 \epsilon_0}{i H_0^2} \R_1(0) \R_2 (0) \R_3(0)  \! \int_{- \infty}^{0} \!\!\!\!\!\! d \tau \Bigl( \frac{3\theta+\eta}{\tau^2} \left[ \R_1 (\tau) \R_2' (\tau) \R_3' (\tau) + sym \right]^* + {\rm c.c.} \nn\\
 -  \frac{\tau\theta' - \eta}{\tau^2}   \left[ \k_2 \cdot \k_3 \,  \R_1 (\tau) \R_2 (\tau) \R_3 (\tau) + sym \right]^* + {\rm c.c.} \Bigr), \qquad \qquad \qquad  \label{Delta-B-R}
\eea
where $\R_i (\tau) \equiv \R (\k_i, \tau)$ is the wave function for comoving curvature perturbations in Fourier space, given by:
\be
 \R_k (\tau) = i \frac{H_0}{2 \sqrt{\epsilon_0 k^3 }} \left(1 +  i k \tau \right) e^{- i k \tau} . \label{R-I}
\ee
In the previous expression, $H_0$ and $\epsilon_0$ correspond to the featureless components of $H$ and $\epsilon$, and may be regarded as constants. Finally, by inserting eq.~(\ref{R-I}) back into eq.~(\ref{Delta-B-R}), and using eq.~(\ref{eta-theta-relation}) to eliminate $\eta$ in favour of $\theta$ we finally obtain~\cite{Mooij:2015cxa}
 \bea
\Delta B &=& \frac{2\pi^4\P_0^2}{(k_1k_2k_3)^3}\int_{-\infty}^{\infty}d\tau \, i \, e^{i(k_1+k_2+k_3)\tau} \nn\\
&& \bigg\{ \frac{6 \theta - \alpha   \tau \theta '  }{2}\bigg[i(k_1k_2k_3)\tau(k_1k_2+k_3k_1+k_2k_3)-(k_1k_2)^2-(k_3k_1)^2-(k_2k_3)^2\bigg]\nn\\
&& \left. -   \frac{ (2 + \alpha ) \theta'}{4\tau}(k_1^2+k_2^2+k_3^2)(1-ik_1\tau)(1-ik_2\tau)(1-ik_3\tau)\right\}, \label{deltaB}
\eea
where $\mathcal P_0 = H_0^2 / 8 \pi^2 \epsilon_0$ is the featureless contribution to the power spectrum. It is useful to parametrize non-Gaussianity with the help of the dimensionless $f_{\rm NL}$-parameter, which may be conveniently defined as:
\be
\label{fnl}
f_{\rm NL} \equiv \frac{10}{3}\frac{k_1k_2k_3}{k_1^3+k_2^3+k_3^3}\frac{(k_1k_2k_3)^2}{(2\pi)^4\P_0^2}\Delta B .
\ee
In the following sections we will use eq.~(\ref{deltaB}) to analyze the presence of features in the primordial bispectrum. Before moving on to that discussion, let us briefly review how eq.~(\ref{general_expression}) is derived. This will allow us to cover the procedure to correlate different configurations of the bispectrum in a much simpler way.

\subsection{Previous work: power spectrum - bispectrum correlation} \label{previous-work}

It is clear that variations of the background quantities $\epsilon$ and $c_s$ will induce the existence of features in both, the power spectrum and bispectrum. The dimensionless power spectrum $\mathcal P(k)$ parametrizes the two-point correlation function of curvature perturbations at the end of inflation as:
\be
\langle \R_{\k} \R_{\k'} \rangle \equiv (2 \pi)^3 \delta(\k + \k') \frac{2 \pi^2}{k^3} \P(k) .
\ee
We may now split the power spectrum into two parts as
\be
\P(k) = \P_0(k) + \Delta \P(k),
\ee
where $ \P_0(k)$ is the piece containing the main featureless contribution, which is determined by the averaged quasi-de Sitter background, and $\Delta \P(k)$ is the piece containing the features which result from the small but rapid variations of the background quantities. By using the in-in formalism to compute both pieces, one deduces:
\be
k^3 \frac{\Delta \mathcal P}{\mathcal P_0 } (k) = - \frac{1 + \alpha}{16} \int_{-\infty}^{0} \!\!\! d \tau \,  \theta''''  \, \sin (2 k \tau) \, . \label{features-delta-P}
\ee
To derive this relation, we have assumed the hierarchy of eq.~(\ref{hierarchy}) and the validity of eq.~(\ref{eta-theta-relation}) introducing the parameter $\alpha$ linking $\eta$ with $\theta$. Notice that the end of inflation happens essentially at $\tau = 0$, which corresponds to $t \to + \infty$. We may now use the trick of extending the domain of integration from $( -Ê\infty ,  \tau )$ to $( -Ê\infty , + \infty )$ by assuming that $\theta$ is odd under the reparametrization $\tau \to - \tau$. This allows us to rewrite eq.~(\ref{features-delta-P}) as:
\be
k^3 \frac{\Delta \mathcal P}{\mathcal P_0 } (k) = - \frac{1 + \alpha}{32 i} \int_{-\infty}^{+\infty} \!\!\! d \tau \,  \theta''''  \, e^{  2 ik \tau} \, . \label{features-delta-P-2}
\ee
This last equation can now be Fourier inverted to obtain a formal expression for $\theta$ in terms of $\Delta \P$. The result is:
\be
\theta  = \frac{1}{1 + \alpha} \frac{2}{\pi i}\int_{-\infty}^{+\infty} \! \frac{d k}{k} \,  \frac{\Delta \mathcal P}{\mathcal P_0 } (k) \, e^{- 2i k \tau} \, . \label{features-delta-P-3}
\ee
Finally, we plug this expression for $\theta$ back into eq.~(\ref{deltaB}). The result of doing this is precisely an equation of the form given in eq.~(\ref{general_expression}). However, because of the hierarchy assumption of eq~(\ref{hierarchy}), the result is strictly valid in the sharp feature limit, so we obtain eq.~(\ref{general_expression_sharp}), with $\beta$ given by
\be
\beta_\alpha (k_1 , k_2 , k_3)  =  \frac{5}{12} \frac{1}{1+\alpha} \frac{k_1 k_2 k_3}{k_1^3 +k_2^3 +k_3^3}  \left[ \alpha +2  \frac{ k_1^2+k_2^2+k_3^2 }{(k_1+k_2+k_3)^2} \right]. \label{beta-alpha}
\ee
In the case $\alpha = 0$, we recover the relation deduced in ref.~\cite{Achucarro:2012fd} valid for features generated as a result of sound speed time variations. On the other hand, if $|\alpha| \to \infty$, we recover the result deduced in ref.~\cite{Palma:2014hra} valid for features that result from rapid variations of the slow roll parameters $\epsilon$ and $\eta$. It is clear that this function $\beta_\alpha$ can handle any configuration function as long as all three $k_i$ are nonzero. In the squeezed limit ($k_1=0,~ k_2=k_3=k$), we get $\beta_\alpha=0$. Then, the largest contribution to $f_{\rm NL}$ comes then from a term proportional to the first derivative of the power spectrum (and independent of the parameter $\alpha$):
\be
f^{\rm (sq)}_{\rm NL} = -\frac{5}{12} \left[ \frac{d}{d \ln k} \frac{\Delta \P}{\P_0}(k) \right]_{k=(k_1 + k_2 + k_3)/2} ,
\ee
consistent with Maldacena's consistency relation. Given the hierarchy of derivatives in the features that we are studying (see eq.~\ref{hierarchy}), we understand that for such sharp features, the squeezed configuration is suppressed with respect to the other configurations: $f_{\rm NL}^{\rm sq}$ involves one less derivative on the power spectrum feature than the non-linearity parameter for any other configuration, such as the equilateral ($k_1=k_2=k_3 = K/3$) and folded ($k_1=k_2= K/4,~ k_3=K/2$) one.

\setcounter{equation}{0}
\section{This work: bispectrum-bispectrum correlation}  \label{corrgen}

Let us now move on to the main computation of this article, namely, finding an expression correlating features at arbitrary non-Gaussian triangle configurations. The procedure will be in fact similar to the one used to correlate the bispectrum with the power spectrum, covered in Section~\ref{previous-work}. To start with, let us parametrize the momenta $k_1$, $k_2$ and $k_3$ defining triangle configurations in the following way:
\be
k_1 = x K,\qquad k_2= y  K,\qquad k_3=z K ,\qquad z\equiv 1-x-y. \label{parametrization-1}
\ee
This parametrization ensures that $k$ is the sum of the three momenta:
\be
K \equiv k_1 + k_2 + k_3 .
\ee
Note that $z$ is not a free parameter, just a shorthand for $(1-x-y)$ that we use to not obscure the symmetry in the problem. In this parametrization, the equilateral configuration is given by the choice $x=y=1/3$, whereas the folded configuration is determined by $x=y=1/4$. For completeness, in the appendix \ref{a1} we summarize our results with a different parametrization, commonly encountered in the literature, in which one fixes one of the momenta (say $k_3$) and uses the ratios $k_1/k_3$ and $k_2/k_3$ as free parameters. Now, inserting the parametrization given in eq.~(\ref{parametrization-1}) back into eq.~(\ref{deltaB}), we obtain
\bea
\Delta B &=& \frac{2 i\pi^4 \P_0^2}{x^3 y^3 z^3 K^7} \int_{-\infty}^{\infty} \!\!\! d\tau \, e^{i K\tau}\nn \\
&& \Bigl[ \frac{6 \theta - \alpha   \tau \theta '  }{2}\left(-[x^2z^2+y^2 z^2+x^2y^2]K^2+i xyz[xz+yz+xy]K^3\tau \right) \nn\\
&& - \frac{ (2 + \alpha ) \theta'}{4\tau} (x^2+y^2+z^2)(1-i K \tau-[xz+yz+xy] K^2\tau^2 +ixyz K^3\tau^3) \Bigr] . \quad\label{dB}
\eea
Since we are interested in sharp features, we will have that $\tau\theta'\gg\theta$. Moreover, the largest contribution comes from the terms highest order in $K$.\footnote{One could perform some integrations by parts to see that terms with a higher number of powers of $K$ are equivalent to terms with a higher number of time-derivatives on $\theta$. That is why focussing on sharp features implies that the highest order terms in $K$ dominate the right hand side of eq.~(\ref{dB}).} Rearranging and performing one partial integration to eliminate $\theta'$ in favor of $\theta$ yields:
\be
(xyz)^2 \frac{K^3}{i\pi^4 \P_0^2}  \Delta B = -\int_{-\infty}^{\infty} \!\!\! d\tau~e^{i K \tau}~\tau^2\theta~\left( x^2+y^2+z^2+\frac{\alpha}{2} \right).
\ee
Now, inverting this expression gives
\be
(xyz)^2 \int_{-\infty}^\infty \!\!\! d K \frac{K^3}{2i\pi^5 \P_0^2}  \Delta B e^{-iK\tau}
=  -\tau^2\theta~\left( x^2+y^2+z^2+\frac{\alpha}{2} \right) , \label{genalpha2}
\ee
so we directly have:
\be
\theta =  \frac{1}{\tau^2} \frac{(xyz)^2}{2[x^2+y^2+z^2]+\alpha } \int_{-\infty}^\infty \!\!\! dK \frac{i K^3}{\pi^5 \P_0^2}  \Delta B e^{-i K \tau}. \label{thetafin2}
\ee
Notice that this result may diverge for specific configurations if the denominator $2[x^2+y^2+z^2]+\alpha$ becomes $0$, which could be the case if $\alpha$ is negative. This only means that for those configurations other terms, that have been neglected to go from eq.~(\ref{dB}) to eq.~(\ref{thetafin2}), will become relevant. For the sake of simplicity, in the present work we omit such terms and focus on those situations where such divergences do not happen. Nevertheless, we should keep in mind that a more accurate treatment would include these omitted terms.

Equation~(\ref{thetafin2}) allows us to compare $\Delta B$ for two different configurations. For a given configuration $1$ we have $k_1=x_1 K$, $k_2 = y_1 K$ and $k_3=z_1 K$, giving us back $\Delta B_1$. For a different configuration $2$ we have $k_1=x_2 K$, $k_2 = y_2 K$ and $k_3=z_2 K$, giving us back $\Delta B_2$. As we already mentioned in the introduction, since the feature takes place at one particular momentum value, the sizes (the sum of the sides) of the two triangles that we are correlating here are equal. That is, we are looking at manifestations of one and the same sharp feature into different configuration functions corresponding to the comoving wavelength of that feature. Now we may use eq.~(\ref{thetafin2}) to obtain two alternative expressions for $\theta$ in terms of $\Delta B_1$ and $\Delta B_2$ respectively. Then, comparing both expressions we obtain a relation between $\Delta B_1$ and $\Delta B_2$ found to be given by:
\bea
\Delta B_1 
&=& \left(\frac{x_2 y_2z_2}{x_1 y_1z_1}\right)^2 ~ \frac{2(x_1^2+y_1^2+z_1^2)+\alpha}{2(x_2^2+y_2^2+z_2^2)+\alpha }  \Delta B_2 . \label{dbfrac}
\eea
We note that we could equally well have worked in terms of $\eta$ rather than $\theta$. Indeed, beginning again from eq.~(\ref{Delta-B-R}) and repeating the same steps leads back to eq.~(\ref{dbfrac}).  To continue, using the definition of $f_{\rm NL}$ given in eq.~(\ref{fnl}) we finally obtain
\bea
f_{\rm NL}^{(1)} 
&=&  \frac{x_1 y_1 z_1}{x_2 y_2 z_2}~\frac{x_2^3+y_2^3+z_2^3}{x_1^3+y_1^3+z_1^3}  \frac{2(x_1^2+y_1^2+z_1^2)+\alpha }{2(x_2^2+y_2^2+z_2^2)+\alpha } f_{\rm NL}^{(2)} ,\label{FINres2}
\eea
which is our main result. For $|\alpha|\gg1$, the last term at the right hand side asymptotes to $1$, and we obtain the relation valid for features generated exclusively by deviations from slow-roll regime through variations of $\epsilon$ and $\eta$. On the other hand, for $\alpha = 0$ we obtain the relation valid for features generated by a variation of the sound speed (parametrizing deviations from canonical inflation).  

Once again we remind the reader that $z$ is not a free parameter, just a shorthand: $z_i\equiv 1-x_i-y_i$ for $i=\{1,2\}$. Inserting $x_1=y_1=1/3$ for the equilateral configuration and $x_2=y_2 = 1/4$ for the folded configuration gives:
\be
\frac{f_{\rm NL}^{\rm eq}}{f_{\rm NL}^{\rm fold}} = \frac{1/27}{1/32}~\frac{5/32}{3/27}~\frac{2/3+\alpha}{3/4+\alpha} = \frac{20}{9}~\frac{2+3\alpha}{3+4\alpha}.  \label{check}
\ee
This result is compatible with the result of our previous work \cite{Mooij:2015cxa}, given in eq.~(\ref{general_expression_sharp}) of this paper. Using the definition for $\beta_\alpha$ in eq.~(\ref{beta-alpha}) that evaluates to
\be
\frac{f_{\rm NL}^{\rm eq}}{f_{\rm NL}^{\rm fold}}  =  \frac{\beta_\alpha^{\rm eq}}{\beta_\alpha^{\rm fold}}= \frac{20}{9}~\frac{2+3\alpha}{3+4\alpha}.
\ee
Therefore, we stress again that the merit and novelty of this section's computation is not its final result, but the fact that it has been derived in a more direct way. It is a pure bispectrum computation, independent of the associated features in the power spectrum. 

In appendix \ref{a1} we will perform the same computation, but now  working in the popular parametrization $x\equiv k_1/k_3$, $y\equiv k_2/k_3$, i.e.
\be
k_1 = x \tilde{k}, \qquad k_2=y \tilde{k}, \qquad k_3=\tilde{k}.
\ee
We verify explicitly that after correct normalization of the triangles in momentum space, we again get to the result in eq.~(\ref{FINres2}).

\setcounter{equation}{0}
\section{Including the squeezed configuration}  \label{corrsq}

The result given in eq.~(\ref{FINres2}) suggests that for the squeezed configuration, the vanishing of one of the momenta leads to a vanishing $f_{\rm NL}^{\rm sq}$. However, that is not quite true. To get to the result in eq.~(\ref{FINres2}), we have considered only the terms that were of highest order in $k$ in our expression in eq.~(\ref{deltaB}) for $\Delta B$. In the squeezed limit, those terms vanish, and we should focus on the highest order in $k$ among the surviving terms. Given the hierarchy of eq.~(\ref{hierarchy}) in which the number of derivatives counts as an order parameter, we expect that $f_{\rm NL}^{\rm sq}$ will come out an order of magnitude smaller than $f_{\rm NL}^{\rm gen}$ for a general non-squeezed configuration. Therefore, let us compare the squeezed configuration
\be
k_1=0 ,\qquad k_2 = K/2, \qquad k_3 = K/2,  \label{sque}
\ee
with the general configuration:
\be
k_1=x   K ,\qquad k_2 = y    K, \qquad k_3 = z  k \equiv (1-x-y) K.  \label{genconfiguration}
\ee
In other words, in the first (squeezed) configuration we have $x_{\rm sq}=0$ and $y_{\rm sq}=z_{\rm sq}=1/2$. The second configuration is general, but we assume that it is far enough from the squeezed limit for the terms of highest power in $K$ in eq.~(\ref{deltaB}) to be the dominant ones.

To proceed, we use eq.~(\ref{deltaB}) directly evaluated with the configuration of eq.~(\ref{sque}), corresponding to the squeezed configuration, and focus on the terms that are highest order in $K$. This leads to:
\be
(x_{\rm sq} y_{\rm sq} z_{\rm sq})^3 K^7 \frac{\Delta B_{\rm sq}(k)}{i\pi^4 \P_0^2} = - \int_{-\infty}^\infty \!\!\!\!\!\! d \tau \, e^{i K \tau} \Bigl[  (6\theta-\alpha\tau\theta') \frac{K^2}{16} +\frac{(2+\alpha)\theta'}{4\tau} \left(1-iK\tau-\frac{K^2\tau^2}{4}\right)   \Bigr]   .
\ee
Here we have again used eq.~(\ref{eta-theta-relation}) to eliminate $\eta$ in favour of $\theta$. The factor $(x_{\rm sq}y_{\rm sq}z_{\rm sq})^3$ (which evaluates to $0$) will be absorbed once we use eq.~(\ref{fnl}) to rewrite $\Delta B_{\rm sq}$ in terms of $f_{\rm NL}^{\rm sq}$. Now, repeating the same steps that we set in the previous section's computation yields an expression for $\theta$ as a function of $\Delta B_{\rm sq}$:
\be
\theta = \frac{4}{\tau^2} \frac{1}{1+\alpha} \frac{(x_{\rm sq} y_{\rm sq} z_{\rm sq})^3}{i\pi^5 \P_0^2} \int_{-\infty}^\infty \!\!\! d K  \, K^3  \left( \frac{\partial}{\partial \ln K} \Delta B_{\rm sq}(K)\right)~ e^{-i K\tau}
\label{thetasq} .
\ee
To obtain this result, we have neglected terms that are subleading with respect to the hierarchy of eq.~(\ref{hierarchy}). Now we can use this expression for $\theta$ to plug it back into eq.~(\ref{dB}) evaluated at the general configuration parametrized in~(\ref{genconfiguration}). Alternatively, we may simply compare eq.~(\ref{thetasq}) with the earlier expression of eq.~(\ref{thetafin2}). We obtain:
\be
\frac{(x y z)^2}{2[x^2+y^2+z^2]+\alpha} \Delta B_{\rm gen}(K) = -\frac{4(x_{\rm sq} y_{\rm sq} z_{\rm sq})^3}{1+\alpha}~\frac{\partial}{\partial \ln K} \Delta B_{\rm sq}(K)  .\label{samestory}
\ee
Then, using the definition for $f_{\rm NL}$ in terms of $\Delta B$ given in eq.~(\ref{fnl}) we finally find:
\be
f_{\rm NL}^{\rm gen}(K) = -\frac{(2 [x^2+y^2+z^2]+\alpha)x y z}{(1+\alpha) (x^3+y^3+z^3)} ~\frac{\partial}{\partial \ln K} f_{\rm NL}^{\rm sq}(K) .  \label{finressq}
\ee
This expression gives us the correlation in momentum space between a squeezed triangle (parametrized in eq.~(\ref{sque})) and any general non-squeezed triangle (parametrized in eq.~(\ref{genconfiguration})) whose sides add up to the same value of $K$. We see that $f_{\rm NL}^{\rm sq}$ is suppressed compared to $f_{\rm NL}^{\rm gen}$. There is a log-derivative in between them, which means a factor of the order parameter for the sharp features that we are studying. Of course, that is in line with what we have found in our previous work \cite{Mooij:2015cxa}: $f_{\rm NL}^{\rm sq}$ is proportional to one derivative less of $\Delta \P$ than $f_{\rm NL}$ for other configurations.

\setcounter{equation}{0}
\section{Discussion and conclusions}

While observations are still fully compatible with canonical single-field slow-roll inflation, they still leave room to study departures in the form of small scale-dependent features. In this work we have continued our study of sharp features, happening within an efold of inflation. \new{Whereas our previous study \cite{Mooij:2015cxa}, was aimed at correlating features in the bispectrum with features in the power spectrum
%, we have learnt that the latter can easily be unobservably small, while the former can still be seen. That is why in this work 
we have now proposed to look at bispectrum-bispectrum correlations, i.e. correlations between different configurations of the momentum triangle. } We have established that if a sharp features show up in some particular momentum space triangle, we can predict its manifestation in any other configuration function. Once observed in one configuration, checking the correlation with other configuration functions will be a very useful tool to find out whether an observed feature can really be explained as resulting from a rapid time variation of the expansion rate $H$ and/or (dependent on the value of $\alpha$) of the speed of sound $c_s$.

In particular, we have indicated that the strength of a non-Gaussian signal (the size of $f_{\rm NL}$) caused by a sharp feature in the inflaton's dynamics depends on the configuration of the momentum triangle. In other words, while changing this configuration (but leaving the sum of the three sides unchanged), the amplitude of the predicted non-Gaussian signal changes as well. In particular, when approaching the squeezed limit, it decreases by an order of magnitude. We feel that this very basic observation might have considerable consequences for analyzing non-Gaussianities. Indeed, when looking for non-Gaussian manifestations of sharp features in the data, one should use templates with a configuration-dependent amplitude.

\subsection*{Acknowledgements }

We would like to thank Ana Ach\'ucarro, \new{Peter Adshead, }Vicente Atal, James Fergusson, Jinn-Ouk Gong, Bin Hu, Subodh Patil, Spyros Sypsas and Jes\'us Torrado for helpful discussions and/or insightful comments on this manuscript. This work was supported by the Fondecyt project number 1130777 (GAP \& AS), by the ``Anillo'' project ACT1122 funded by the ``Programa de Investigaci\'on Asociativa" (GAP \& GP), by the Fondecyt 2015 Postdoctoral Grant 3150126 (SM), and by the CONICYT-PCHA/MagisterNacional/2013-221320624 (AS).

\begin{appendix}

\renewcommand{\theequation}{\Alph{section}.\arabic{equation}}
\setcounter{equation}{0}
\section{Alternative parametrization}

In this Appendix we want to repeat the main text's computation in a different parametrization of the momentum space triangles under consideration. Now the ratios $k_1/k_3$ and $k_2/k_3$ are our free parameters. We feel that the parametrization used in the main text is somewhat more transparent, but given the popularity of this other parametrization, this Appendix might be useful for comparing our results to the literature.

\subsection{General configurations} \label{a1}

We consider a general configuration function:
\be
k_1 = x k,\qquad k_2= y  k,\qquad k_3= k.
\ee
So for $x=y=1$ we should recover the equilateral results, and for $x=y=1/2$ we are back to the folded case. Inserting eq.~(\ref{deltaB}) for our general configuration gives (still in terms of both $\theta$ and $\eta$):
\bea
\frac{x^3y^3}{2}\frac{k^7}{i\pi^4 \P_0^2}  \Delta B_{\rm gen} &=& \int_{-\infty}^{\infty} d\tau e^{i[x+y+1] k\tau}\Bigl[ (3\theta+\eta)\left(-[x^2+y^2+x^2y^2]k^2+i xy[x+y+xy]k^3\tau \right) \nn\\
&& \qquad +\frac{\eta-\tau\theta'}{\tau^2}\frac{1+x^2+y^2}{2}(1-i[x+y+1]k\tau-[x+y+xy] k^2\tau^2 +ixy k^3\tau^3) \Bigr] . \nn\\
\eea
We focus on the sharpest terms (the ones proportional to $k^3$) and perform three partial integrations, yielding three factors of $([x+y+1]k\tau)^{-1}$. We obtain:
\bea
\frac{x^3y^3}{2}\frac{k^7}{i\pi^4 \P_0^2}  \Delta B_{\rm gen} 
 &=& \int_{-\infty}^{\infty} d\tau \frac{e^{i[x+y+1] k\tau}}{2[x+y+1]^3}\Bigl[ \eta''' xy (1+x+y)^2  \nn\\
&& \qquad \qquad\qquad \qquad \qquad - \tau \theta'''' xy  \left[1+x^2+y^2\right]\Bigr]\tau.
\eea
Inverting this relation gives:
\be
\frac{x^2y^2(x+y+1)^4}{2}  \int_{-\infty}^\infty dk  \frac{k^7}{i\pi^5 \P_0^2}  \Delta B_{\rm gen} e^{-i[x+y+1]k\tau} 
=
\Bigl[ \eta''' (1+x+y)^2 - \tau \theta''''  \left[1+x^2+y^2\right]\Bigr]\tau.
\ee
Now we use our relation $\eta=-\frac{\alpha}{2} \tau \theta'$. That gives:
\bea
\frac{x^2y^2(x+y+1)^4}{2}  \int_{-\infty}^\infty dk  \frac{k^7}{i\pi^5 \P_0^2}  \Delta B_{\rm gen} e^{-i[x+y+1]k\tau} &=& -\left( 1+x^2+y^2+\frac{\alpha}{2}(1+x+y)^2   \right)\theta''''\tau^2 \nn\\
&=& \left( (1+x+y)^2 +\frac{2}{\alpha} \left[ 1+x^2+y^2 \right]  \right) \eta''' 
\tau. \nn\\ \label{genalpha}
\eea
Now we can express $\theta$ and $\eta$ as functions of $\Delta B$. Isolating $\theta$ and performing four integrations with respect to $\tau$ (assuming that the $\tau$-derivative's dominant effect is on $\theta$, and not on the factors of $\tau$) gives
\be
\theta = - \frac{1}{\tau^2} \frac{x^2 y^2}{2+2x^2+2y^2+\alpha(1+x+y)^2 }  \int_{-\infty}^\infty dk  \frac{k^3}{i\pi^5 \P_0^2}  \Delta B_{\rm gen} e^{-i[x+y+1]k\tau},  \label{thetafin}
\ee
while for $\eta$ we get:
\be
\eta =-\frac{1}{\tau}  \frac{\alpha x^2 y^2 (x+y+1)}{2\left( 2 \left[ 1+x^2+y^2 \right] + \alpha(1+x+y)^2  \right)}  \int_{-\infty}^\infty dk  \frac{k^4}{\pi^5 \P_0^2}  \Delta B_{\rm gen} e^{-i[x+y+1]k\tau}. \label{etafin}
\ee
We can now turn to the computation of the ratio of the values for $f_{\rm NL}$ for two different configurations. For configuration $1$ we have $k_1=x_1 k^{(1)}$, $k_2 = y_1 k^{(2)}$ and $k_3=k^{(1)}$. For  configuration $2$ we have $k_1=x_2 k^{(2)}$, $k_2 = y_2 k^{(2)}$ and $k_3=k^{(2)}$. Normalization of the triangles requires
\be
(x_1+y_1+1) k^{(1)} = (x_2+y_2+1) k^{(2)}.  \label{normtri}
\ee
(Note that the advantage of the parametrization used in the main text is that this normalization is automatically taken care of.) Using the definition of $f_{\rm NL}$ given in eq.~(\ref{fnl}) expresses the ratio of the two $f_{\rm NL}$-values as a ratio of the two bispectrum perturbations:
\be
\frac{f_{\rm NL}^{(1)}}{f_{\rm NL}^{(2)}} = \frac{ (x_1 y_1)^3 (x_2^3+y_2^3+1)}{ (x_2 y_2)^3(x_1^3+y_1^3+1)} \left(\frac{k^{(1)}}{k^{(2)}}  \right)^6 \frac{\Delta B_1}{\Delta B_2}.  \label{fnlproto}
\ee
The ratio of values for $\Delta B$ for the two configurations follows from our result eq.~(\ref{thetafin}) for $\theta$. Formally expressing $\theta$ as a function of $\Delta B_1$ and as a function of $\Delta B_2$ and comparing these two gives:
\bea
\frac{\Delta B_1}{\Delta B_2}
&=&\left(\frac{x_1+y_1+1}{x_2+y_2+1} \right)^4   \left(\frac{x_2 y_2}{x_1 y_1}\right)^2   \frac{2(1+x_1^2+y_1^2)+\alpha(1+x_1+y_1)^2}{2(1+x_2^2+y_2^2)+\alpha(1+x_2+y_2)^2 }.
\eea
Of course, expressing $\eta$ as a function of $\Delta B_1$ and as a function of $\Delta B_2$, and comparing these two expressions gives the same result.

Finally we turn to the ratio of the values for $f_{\rm NL}$ that we had found in eq.~(\ref{fnlproto})
\bea
\frac{f_{\rm NL}^{(1)}}{f_{\rm NL}^{(2)}} & = & \left(\frac{x_1 y_1}{x_2 y_2}\right)^3   \frac{x_2^3+y_2^3+1}{x_1^3+y_1^3+1}  \left( \frac{x_2+y_2+1}{x_1+y_1+1}\right)^6  \left(\frac{x_1+y_1+1}{x_2+y_2+1} \right)^4   \left(\frac{x_2 y_2}{x_1 y_1}\right)^2  \nn\\
&& \qquad \qquad \qquad \times \frac{2(1+x_1^2+y_1^2)+\alpha(1+x_1+y_1)^2 }{2(1+x_2^2+y_2^2)+\alpha(1+x_2+y_2)^2 }\nn\\
&=&  \frac{x_1 y_1}{x_2 y_2} \frac{x_2^3+y_2^3+1}{x_1^3+y_1^3+1}  \left( \frac{x_2+y_2+1}{x_1+y_1+1}\right)^2    \frac{2(1+x_1^2+y_1^2)+\alpha(1+x_1+y_1)^2 }{2(1+x_2^2+y_2^2)+\alpha(1+x_2+y_2)^2 } , \label{FINres}
\eea
and that is the final result. Inserting $x_1=y_1=1$ for the equilateral configuration and $x_2=y_2 = 1/2$ for the folded configuration gives
\be
\frac{f_{\rm NL}^{\rm eq}}{f_{\rm NL}^{\rm fold}} = \frac{1}{1/4} \frac{5/4}{3} \left(\frac{2}{3}\right)^2 \frac{6+9\alpha}{3+4\alpha} = \frac{20}{9} \frac{2+3\alpha}{3+4\alpha},  
\ee
as we had already found in eq.~(\ref{check}).

\subsection{Squeezed configuration} \label{a2}

We compare the squeezed configuration
\be
k_1=0 ,\qquad k_2 = k, \qquad k_3 = k,
\ee
with the general configuration:
\be
k_1=x  \tk ,\qquad k_2 = y   \tk, \qquad k_3 = \tk.
\ee
In other words, in the first configuration we have $x_{\rm sq}=0$ and $y_{\rm sq}=1$. The normalization is as in eq.~(\ref{normtri}):
\be
2 k =(x+y+1) \tk.  \label{normsq}
\ee
Inserting eq.~(\ref{deltaB}) we lose the terms that are highest order in $k$. We get:
\be
(x_{\rm sq} y_{\rm sq})^3 k^7 \frac{\Delta B_{\rm sq}(k)}{2i\pi^4 \P_0^2} = \int_{-\infty}^\infty d \tau   e^{2i k \tau} \Bigl[  -(3\theta+\eta) k^2 +\frac{\eta-\tau\theta'}{\tau^2}  (1-2ik\tau-k^2\tau^2)   \Bigr]   .
\ee
This yields for $\eta$ and $\theta$ 
\bea
\theta &= &\frac{1}{\tau}  \frac{1}{1+\alpha} (x_{\rm sq} y_{\rm sq})^3   \int_{-\infty}^\infty d k   k^4 \frac{\Delta B_{\rm sq}(k)}{4\pi^5 \P_0^2}    e^{-2i k\tau}\nn\\
&=& \frac{1}{\tau^2}  \frac{1}{1+\alpha} (x_{\rm sq} y_{\rm sq})^3   \int_{-\infty}^\infty d k   k^3  \left( \frac{\partial}{\partial \ln k} \frac{\Delta B_{\rm sq}(k)}{8i\pi^5 \P_0^2}\right)  e^{-2i k\tau}
\label{thetasq2}	
\eea
and
\bea
\eta& =& - \frac{\alpha}{1+\alpha} (x_{\rm sq} y_{\rm sq})^3 \int_{-\infty}^\infty d k   k^5  \frac{\Delta B_{\rm sq}(k)}{4i\pi^5 \P_0^2}    e^{-2i k\tau}\nn\\
&=& \frac{1}{\tau}  \frac{\alpha}{1+\alpha} (x_{\rm sq} y_{\rm sq})^3 \int_{-\infty}^\infty d k   k^4 \left( \frac{\partial}{\partial \ln k} \frac{\Delta B_{\rm sq}(k)}{8\pi^5 \P_0^2}\right)   e^{-2i k\tau}
 \label{etasq2}.
\eea
Here we have rewritten $\theta$ and $\eta$ such that we can compare them with our earlier expressions eq.~(\ref{thetafin}). Indeed, comparing eq.~(ref{thetafin}) and eq.~(\ref{thetasq2}) leaves us with (upon imposing the normalization given in eq.~(\ref{normsq})):
\be
f(\alpha,x ,y )  \left(\frac{2}{x +y +1}\right)^4 (x y)^3  \Delta B_{\rm gen}(\tk) = -\frac{1}{1+\alpha} \frac{1}{8} (x_{\rm sq} y_{\rm sq})^3 \frac{\partial}{\partial \ln k} \Delta B_{\rm sq}(k), 
\ee
where we have introduced the function $f(\alpha,x,y)$:
\be
f(\alpha,x,y) = \frac{1}{x y}  \frac{1}{2(1+x^2+y^2)+\alpha(1+x+y)^2 } .
\ee
The same result follows, of course, from comparing the expressions eq.~(\ref{etafin}) and eq.~(\ref{etasq2}) for $\eta$.  Finally, inserting equation eq.~(\ref{fnl}) now gives us
\be
f(\alpha,x,y)    \left(\frac{2}{x+y+1}\right)^4 \left(x^3+y^3+1\right) \frac{1}{\tk^6}  f_{\rm NL}^{\rm gen}(\tk) = -\frac{1}{1+\alpha}  \frac{1}{4} \frac{1}{k^6} \frac{\partial}{\partial \ln k} f_{\rm NL}^{\rm sq}(k),
\ee
and upon using the normalization eq.~(\ref{normsq}) that gives
\be
f(\alpha,x,y)    \left(x+y+1\right)^2 \left(x^3+y^3+1\right)  f_{\rm NL}^{\rm gen}(\tk) = -\frac{1}{1+\alpha}  \frac{\partial}{\partial \ln k} f_{\rm NL}^{\rm sq}(k),
\ee
which is indeed consistent with our result in eq.~(\ref{finressq}). (Note that since $f_{\rm NL}$ is a scale-independent parameter, it is independent of the used parametrization.)

\end{appendix}

\end{document}